# AN INFINITE FRACTAL COSMOS

Robert L. Oldershaw

12 Emily Lane

Amherst, MA 01002

rloldershaw@amherst.edu

7 pages; 0 figures; 0 Tables

**ABSTRACT:**

The paradigmatic transition from a small finite universe to an infinite unbounded fractal cosmos is briefly put into historical context and discussed.

"An Infinite Fractal Cosmos"

With regard to cosmology we live at a very privileged time. Previously, when we learned about bold revolutionary paradigm changes, in which our fundamental ideas about nature underwent radical revisions, the drama had already taken place in the distant past. Today, however, we have the rare opportunity of witnessing at first hand a profound transformation in our basic understanding of how nature is structured. This ongoing change of cosmological paradigms from a "small" finite cosmos to an infinite fractal cosmos began about three decades ago and is in full swing at present.

The short version of what is happening is as follows. For over half a century the Big Bang model of the Universe has provided an adequate explanation for the basic cosmological observations: a global expansion, an approximately uniform background of microwave radiation and a unique set of atomic element abundances. However, there were some technical problems with this model, such as an *acausal* beginning of spacetime, an unexpected failure to observe magnetic monopoles, a surprisingly high degree of large-scale uniformity, and an enigmatic knife-edge balance between open (infinite) and closed (finite) states. Then in the early 1980s Alan Guth showed how these and other problems with the Big Bang model could be solved in one fell swoop with the Inflationary Scenario (Guth and Steinhardt, 1984), which postulated a very brief period of ultra-rapid expansion shortly after the Big Bang. The theory of Inflation gained analytical and observational support over time and it is now fully accepted by cosmologists as a cornerstone of the Big Bang paradigm. But, an ironic thing has happened. Although the Inflationary Scenario was developed to rescue the Big Bang model, the most logical consequence of pursuing the concept of Inflation is the replacement of the Big Bang Paradigm with a much grander and more encompassing paradigm.

According to Guth (1997) and a growing number of leading cosmologists, the most natural version of Inflation theory is Eternal Inflation, in which Inflation is, was and always will be occurring on an infinite number of size scales. The new general paradigm that cosmologists and natural philosophers have arrived at by several different routes is an infinite fractal hierarchy that has "universes" within "universes" without end. The astronomer Carl Sagan (1980) once referred to the fundamental concept of an infinite fractal universe as "strange, haunting, evocative – one of the most exquisite conjectures in science or religion."

That is the basic story, but because of the profound impact that the fractal paradigm will have on our understanding of the Universe and our place within that Universe, it is important to explore the implications of this new understanding of nature. Firstly, there is no edge or boundary of the Universe; space is infinite in all directions. What we used to refer to as "The Universe" can be more appropriately called the "observable universe" (note the small "u") or the "Hubble Volume," and it is only a tiny part of what one might call our local metagalaxy or "level 1 universe." We currently have no way of determining the total size of our metagalaxy or the number of galaxies that compose it, but we could reasonably assume that both figures would be vastly beyond anything previously contemplated. There would be an infinite number of these metagalaxies, and on an unimaginably large scale they would be organized into "level 2 universes", and so on without limit.

Secondly, time is also infinite in the unbounded fractal universe.  Whereas our Hubble Volume may have come into being and started to expand approximately 13.7 billion years ago, the Universe has always existed and always will.  *Parts* of the Universe may be created or annihilated, or may undergo expansion or contraction, but overall the infinite fractal hierarchy remains eternal and unchanged *as a whole*, and therefore it is without temporal limits.

Thirdly, there is no limit to size scales.   In the infinite fractal paradigm there is no class of largest objects that would cap off the cosmological hierarchy; the hierarchy is infinite in scale.  This fact removes one of the more suspect aspects of the old paradigm.  Natural philosophers had long noted the discomforting fact that within the context of the Big Bang paradigm humans found themselves roughly in the *middle* of nature's hierarchy of size scales.  This anthropocentric state of affairs seemed to violate the Copernican principle, which asserts that humans are *not* at the center of the universe.  When humans do appear to be at the center of things, we should strongly suspect that some form of bias is leading us astray.  In an infinite fractal hierarchy there is no center of the Universe whatsoever, nor is there <u>any</u> preferred reference frame or scale.

Some interesting questions immediately arise.  Why are fractal hierarchies so ubiquitous in nature?  By studying empirical phenomena within the observable universe, how much will we be able to learn scientifically about the parts of the Universe that lie beyond our observational limits?  Does the infinite cosmological hierarchy have a bottom-most scale of subatomic particles, as is currently assumed, or is this another artificial boundary in an infinite fractal cosmos that actually extends without limits to ever-smaller scales?

Centuries ago Immanuel Kant, J. H. Lambert, Spinoza and a few others proposed an infinite hierarchical model of the universe based largely on natural philosophy.  This general hierarchical paradigm never garnered a large following, but like the legendary Phoenix it kept arising from the ashes of neglect.  In the 1800s and 1900s quite a few scientists, including E. E. Fournier d'Albe, F. Selety, C. V. L. Charlier and G. de Vaucouleurs, argued for hierarchical cosmological models based on the hierarchical organization within the observable universe.  Then towards the end of the 1970s, the mathematician B. B. Mandelbrot (1977, 1983) gave the hierarchical paradigm new life and widespread exposure by developing the mathematics of fractal geometry and demonstrating that fractal phenomena based on hierarchical self-similarity are ubiquitous in nature.  In this way natural philosophers, research scientists, mathematicians and now theoretical physicists have all found their way, slowly but surely, to the infinite fractal paradigm.  There are many routes to this paradigm, and certainly there are a large number of distinct versions (Oldershaw, 2001; Nottale, 1993; Tegmark, 2003; Baryshev and Teerikoorpi, 2003; Baryshev, 2008) of the basic paradigm, each with its own unique theoretical explanations for why and how nature is organized in this manner.  For example, the present author has shown how a further generalization of General Relativity involving discrete self-similar scaling of the interactions between matter and the geometry of the spacetime manifold leads to an unbounded discrete self-similar cosmos (Oldershaw, 2007).

Although we do not yet know how all of the technical details of the fractal cosmos will be resolved, the general paradigm that nature is an infinite hierarchy of worlds within worlds has finally arrived, and it is likely to be our dominant paradigm for the foreseeable future.